\documentclass[aps,prl,twocolumn,floatfix,showpacs,noshowkeys]{revtex4}

\usepackage{amsmath,amssymb,bm,color,graphicx}
\newcommand{\ket}[1]{|{#1} \rangle}

\begin{document}
\preprint{Alzhar Uni. Math. Dep.}
\title{Some entanglement features of three-atoms Tavis-Cummings model:\\ Cooperative case}
\author{M. Youssef}
\email{mohamadmath@yahoo.com}

\affiliation{Department of Mathematics, Faculty of Science, Al-Azhar
University, Nasr City, Cairo 11884, Egypt}
\author{N. Metwally}
\affiliation{Department, Faculty of Science, South Valley
University, Aswan, Egypt}
\author{A.-S. F. Obada}
\affiliation{Department of Mathematics, Faculty of Science, Al-Azhar
University, Nasr City, Cairo 11884, Egypt}

\date{\today}
\begin{abstract}
In this paper we consider a system of identical three two-level
atoms interacting at resonance with a single-mode of the quantized
field in a lossless cavity. The initial cavity field is prepared in
the coherent state while the atoms are taken initially to be either
in the uppermost excited state "$|eee\rangle$" or The
$\textmd{GHZ}$-state or the $\textmd{W}$-state. For this system we
investigate different kinds of atomic inversion and entanglement,
which arise between the different parts of the system due to the
interaction. Also the relationship, between entanglement and some
other nonclassical effects in the statistical properties, such as
collapses and revivals in the atomic inversion where superharmonic
effects appear, is discussed. The $Q$-functions for different cases
are discussed. Most remarkably it is found that the
$\textmd{GHZ}$-state is more robust against energy losses, showing
almost coherent trapping and Schr\"{o}dinger-cat states can not be
produced from such state. Also the entanglement of
$\textmd{GHZ}$-state is more robust than the $\textmd{W}$-state.
Another interesting feature found is that the state which has no
pairwise entanglement initially will have a much improvement of such
pairwise entanglement through the evolution. Sudden death and sudden
revival of atoms-pairwise entanglement are produced with the
$\textmd{W}$-state.
\end{abstract}



 \maketitle
\section{Introduction}
The quantum entanglement phenomenon is not only one of the most
interesting features of the quantum theory \cite{Peresbook}, that
signifies it from the classical theory, but also lies at the heart
of the new rapidly developing area known as the quantum information
processing \cite{Niels}. It is one of the crucial resources required
in the applications in this new area of science, which include,
quantum computation\cite{Deutsch}, quantum teleportation
\cite{been1}, quantum dense coding \cite{Been2} and quantum
cryptography \cite{hill}. In quantum optics domain, the interaction
of an atom with a quantized electromagnetic field mode described by
the Jayens-Cumming model\cite{jc,shor}leads to an entanglement of
these two systems such that the total state vector cannot be written
as a product of the time-dependent atomic and field component
vectors \cite{Phonx1,Gea1,Gea2}. To quantify entangled states, one
should know whether they are pure or mixed states. Thus, if the
entangled state is in a pure state, then it is sufficient to use von
Neumann entropy as a measure of entanglement. Many efforts have been
devoted to quantify entanglement, particularly for mixed states of a
bipartite system, and a number of measures have been proposed, such
as entanglement of formation, relative entropy of entanglement and
negativity. The Peres-Horodecki criterion for separability
\cite{peres,horo0} leads to a natural computable measure of
entanglement, called negativity \cite{zycz1,zycz2,vidal}. It has
been proved that the negativity ${\mathcal{N}}(\rho )$ is an
entanglement monotone and therefore can be used as a good measure of
entanglement \cite{vidal}.\\

  On the other hand the cooperative nature of the interaction of a
quantized radiation field with a system of two-level atoms has first
been treated by Dicke \cite{Dick}. A particular case of the Dicke
model, when the atoms interact with a single-mode radiation field
inside a cavity was investigated by Tavis and Cummings TCM
\cite{Tcm1,Tcm2}. The interaction between a field and an ensemble of
atoms develop correlations between the field and the atomic systems
and between the atomic systems parties as well in the course of
their dynamics. Quantifying this quantum correlations(entanglement)
is one of the open problems. Multi-qubit systems are of interest to
be investigated both theoretically and experimentally. In recent
years, great achievements have been made in the application of
three-qubit states, because understanding of entanglement and
dynamics of three-qubit system is an important matter, for example,
it has many applications in quantum cryptography \cite{Julia},
quantum
computation \cite{Pedersen} and quantum gates \cite{monz,joshi}.\\

  In this paper, we consider a system of identical three two-level
atoms interacting at resonance with a single-mode of the quantized
field in a lossless cavity. The initial cavity field is prepared in
the coherent state while the atoms may assume different initial
states. For this system, we investigate some kinds of entanglement
which arise between the different parts of the system due to the
interaction. We use the concurrence, the generalized I-concurrence
and the negativity to quantify entanglement between different
parties as well as the 3-particle residual entanglement. Also we
look at the relationship between this entanglement and some other
nonclassical effects in the statistical properties such as collapses
and revivals in the atomic inversion. The field dynamics and
atoms-field entanglement are discussed through $Q$-function.\\

  This paper is organized as follows. In section 2 we
introduce the system and its solution. Section 3 is devoted to the
atomic inversions. In section 4 we study the field atom
entanglement, two-atom entangle and the atomic system purity.
Section 5 is devoted to the residual 3-particle entanglement. The
field dynamics in phase space, the possibility of having a cat state
and the atoms-field entanglement are addressed in section 6. Finally
in section 7. we conclude the paper with a discussion of the
results.
\maketitle
\section{The model and its time evolution}
 We consider a system of three identical two-level atoms interacting
with a quantized single-mode electromagnetic field. Under the
rotating-wave approximation and resonant condition, the Hamiltonian
of this system reads:
\begin{eqnarray}\label{ham}
\hat{H}&=&\hat{H}_{0}+\hat{H}_{I}=\omega_{f}\left(\hat{a}^{\dagger
}\hat{a}+\frac{1}{2}\sum_{i=a,b,c}\hat{\sigma}_{z}^{(i)}\right)\nonumber\\
&+&g\left(\hat{a}\sum_{i=a,b,c}\hat{\sigma}_{+}^{(i)}+
\hat{a}^{\dagger}\sum_{i=a,b,c}\hat{\sigma}_{-}^{(i)}\right)
\end{eqnarray}
where ($\hbar=1$). The terms $\hat{H}_{0}$ and $\hat{H}_{I}$
represent the free and interaction hamiltonians respectively;
$\omega _{f}$ is the field frequency, and equals the atomic
transition frequency on resonance; $\hat{\sigma}_{+}^{(i)}$,
$\hat{\sigma}_{-}^{(i)}$ and $\sigma_{z}^{(i)}$ are the the usual
raising,lowering and inversion operators for the i$^{th}$ atom,
satisfying
$[\hat{\sigma}_{+}^{(i)},\hat{\sigma}_{-}^{(i)}]=\sigma_{z}^{(i)}$,
$[\hat{\sigma}_{z}^{(i)},\hat{\sigma}_{\pm}^{(i)}]=\pm2\sigma_{\pm}^{(i)}$,
$[\hat{\sigma}_{\mu}^{(i)},\hat{\sigma}_{\nu}^{(j)}]=0$ while
$\hat{a}^{\dagger}(\hat{a})$ is the Bose creation (annihilation)
operator for the quantized field mode satisfying the commutation
relations $[\hat{a},\hat{a}^{\dag}]=1$, and $g$ \ is the coupling
constant. Since $[\hat{H}_{0},\hat{H}_{I}]=0$, it follows that, the
Hamiltonian (\ref{ham}) conserves the total number of excitations
$N$, i.e. the total excitation operator
\begin{equation}
\hat{N}=\hat{a}^{\dagger }\hat{a}+\frac{1}{2}\,\sum_{i=a,b,c}\text{
}\hat{\sigma}^{(i)}_{z}
\end{equation}
is a constant of motion. This provides a decomposition for the
system Hilbert space as ${\mathcal H}=$
$\sum_{n=0}^{\infty}\oplus{\mathcal H}_{n}$ such that, ${\mathcal
H}_{0}=\{|g,g,g;0\rangle\}$, ${\mathcal H}_{1}=\{|g,g,g;1\rangle,
|g,e,g;0\rangle, |g,g,e;0\rangle, |e,g,g;0\rangle\}$ and ${\mathcal
H}_{2}=$ $\{|g,g,g;2\rangle, |e,g,g;1\rangle, |g,e,g;1\rangle,
|g,g,e;1\rangle, |e,e,g;0\rangle$ $, |e,g,e;0\rangle,
|g,e,e;0\rangle\}$ are a one-dimensional , 4-dimensional and
7-dimensional eigensubspaces of $N$ and
${\mathcal H}_{n+3}|_{n=0}^{\infty}=\{|eee;n\rangle,
|eeg;n+1\rangle, |ege;n+1\rangle,$ $ |gee;n+1\rangle,
|egg;n+2\rangle ,|geg;n+2\rangle,|gge;n+2\rangle, |ggg;n+3\rangle
\}$

are the eight-dimensional eigensubspaces of $N$. In the Hilbert
space constituted by the above basis, the interaction $\hat{H}_{I}$
is a diagonal blocks matrix made up of $8\times 8$ sub-matrices,
every sub-matrix represents a subspace corresponding to a definite
excitation number $N$. However we found that it is more appropriate
for this system to use the Dicke states \cite{Dick} as a basis,
because of the degeneracy produced by the symmetry. For $N$
spin-$1/2$ particle system, the Dicke states are defined as the
states $|S,m_{s}\rangle$ that are common eigenstates of both the
square of the total spin operator $\hat{S}^{2}$ and its component
along $z$-axis "the quantization axis " $\hat{S_{z}}$ with the
corresponding eigenvalues $S(S+1)\hbar$ and $m_{s}\hbar$. For our
system the Dicke states in terms of the above mentioned product
states are given by
\begin{eqnarray}\label{Dick}
|{\cal D}_{1}\rangle&=&|eee;n\rangle\nonumber\\
|{\cal D}_{2}\rangle&=&\frac{1}{\sqrt{3}}\left
(|eeg;n+1\rangle+|ege;n+1\rangle+
|gee;n+1\rangle\right)\nonumber\\
|{\cal
D}_{3}\rangle&=&\frac{1}{\sqrt{3}}\left(|egg;n+2\rangle+|geg;n+2\rangle+
|gge;n+2\rangle\right)\nonumber\\
|{\cal D}_{4}\rangle&=&|ggg;n+3\rangle\nonumber\\
|{\cal
D}_{5}\rangle&=&\frac{1}{\sqrt{6}}\left(2|gee;n+1\rangle-|ege;n+1\rangle-
|eeg;n+1\rangle\right)\nonumber\\
|{\cal
D}_{6}\rangle&=&\frac{1}{\sqrt{2}}\left(|eeg;n+1\rangle-|ege;n+1\rangle\right)
\nonumber\\
|{\cal
D}_{7}\rangle&=&\frac{1}{\sqrt{6}}(|geg;n+2\rangle+|gge;n+2\rangle-2
|egg;n+2\rangle)\nonumber\\
|{\cal
D}_{8}\rangle&=&\frac{1}{\sqrt{2}}(|gge;n+2\rangle-|geg;n+2\rangle )
\end{eqnarray}
The first four states are fully symmetric with respect to the
permutation of particles, while the other four states are in fact a
two state of mixed symmetry, which correspond to two degenerate
representations, each of dimension 2 \cite{Rai}. By substituting
into the Schr\"{o}dinger equation
$-i\frac{d|\Psi(\tau)\rangle}{d\tau}=\hat{H}_{I}|\Psi(\tau)\rangle$,
where $\tau=gt$, then the wave function generally can be written as
$|\Psi(\tau)\rangle=\sum\limits_{i=1}^{8}X_{i}^{(n)}(\tau)|{\cal
D}_{i}\rangle$. We found, as expected, that the equations of the
probability amplitudes which correspond to the two degenerate states
are decoupled from each other and from the four amplitudes which
correspond to the fully symmetric Dicke states. The initial state of
the system will be taken as follows:
\begin{eqnarray}
|\Psi (0)\rangle &=&|\Psi_{a}(0)\rangle\otimes |\Psi
_{f}(0)\rangle\nonumber\\
|\Psi_{a}(0)\rangle&=&C_{e}|e,e,e\rangle+\frac{1}{\sqrt{3}}C_{w1}(|e,e,g\rangle
+|e,g,e\rangle+|g,e,e\rangle)\nonumber\\&+&\frac{1}{\sqrt{3}}C_{w2}(|g,g,e\rangle
+|g,e,g\rangle+|e,g,g\rangle)+C_{g}|g,g,g\rangle\nonumber\\
|\Psi_{f}(0)\rangle&=&\sum_{n=0}^{\infty }q_{n}|n\rangle;\quad
q_{n}= e^{(-|\alpha_{0}|^{2}/2\,)}\frac{\alpha_{0}^n}{\sqrt{n!}}
\end{eqnarray}
where $|C_{e}|^2+|C_{w1}|^2+|C_{w2}|^2+|C_{g}|^2=1$. For the above
initial atomic states the wave function evolve only inside the
subspace spanned by the fully symmetric Dicke states and the
probability amplitudes of the other state vanishes, hence our system
is equivalent to a four-level Dicke atom. This is a generalization
to the case of two identical atoms \cite{kudr}, where the
antisymmetric states do not participate in the dynamics. Solving the
coupled four equations of the probability amplitudes we get the wave
function of system at any time
\begin{eqnarray}\label{Amp}
|\Psi(\tau)\rangle &=&\sum_{n=0}^{\infty}\sum_{i=1}^{4}X_{i}^{(n)}(\tau)|{\cal D}_{i}\rangle\nonumber\\
\textmd{X}(\tau)&=&[X_{1}^{(n)}(\tau),X_{2}^{(n)}
(\tau),X_{3}^{(n)}(\tau),X_{4}^{(n)}(\tau)]^{\textmd{T}}\nonumber\\
&=&{\cal U}\textmd{X(0)}
\end{eqnarray}
where the evolution matrix ${\cal U}$ is given by
\begin{equation}\label{evm}
{\cal U}={\begin{pmatrix}
{\cal U}_{11} & {\cal U}_{12}&{\cal U}_{13}&{\cal U}_{14}\\
{\cal U}_{21} & {\cal U}_{22}&{\cal U}_{23}&{\cal U}_{24}\\
{\cal U}_{31} & {\cal U}_{32}&{\cal U}_{33}&{\cal U}_{34}\\
{\cal U}_{41} & {\cal U}_{42}&{\cal U}_{43}&{\cal U}_{44}\\
\end{pmatrix}}
\end{equation}
The explicit form of the matrix elements are given in the appendix,
where in there we have
\begin{eqnarray}\label{Rabi}
\mu_{1,2}&=&\frac{1}{2}\left(\delta\pm\sqrt{\delta
^{2}-36\eta ^{2}\gamma^{2}}\right)\nonumber \\
\delta&=&(4\beta^{2}+3\gamma^{2}+3\eta ^{2})\nonumber\\
\gamma &=&\sqrt{n+1},\quad \beta =\sqrt{n+2},\quad \eta =\sqrt{n+3}
\end{eqnarray}
Note that $X_{i}^{(n)}$ has $q_{n+i-1}$ as a factor, where
i=1,2,3,4. By using this wave function, we discuss different aspects
of the system in what follows.
\maketitle
\section{Atomic inversions}
\begin{figure*}
\centering
\includegraphics[width=\textwidth]{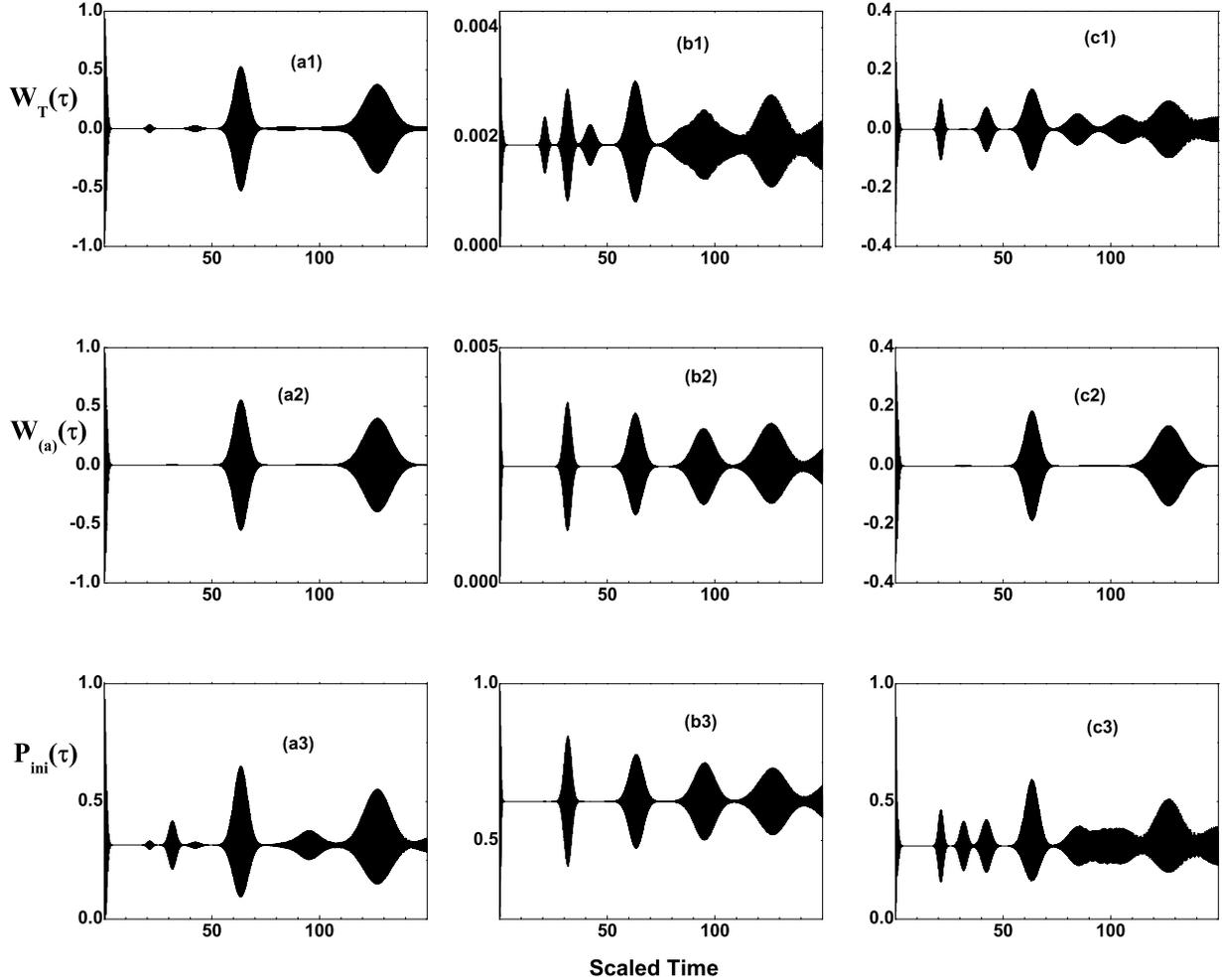}
\caption{Evolution of the total atomic inversions $W_{T}(\tau)$,
single-atom inversions $W_{a}(\tau)$ and the initial state
populations $P_{ini}(\tau)$, Eq.(\ref{inv}) for different initial
atomic states (a)$|eee\rangle$-initial state,(b)
$|\textmd{GHZ}\rangle$-initial state and (c)
$|\textmd{W}\rangle$-initial state. The field in a coherent state
with $|\alpha|^{2}=100$}\label{P1_inv}
\end{figure*}
Energy is the primary quantity determining the properties of
physical systems. The atomic inversion and levels occupation
probabilities are the simplest nontrivial physical quantities in the
atom-field interaction, that display the exchange of energy between
the field and the atoms. More important to our investigation, the
atomic inversions display the primary nonclassical effect namely,
the collapse and revival structure \cite{Eber1,Eber2,Eber3,Eber4}
from which we can have information about the atoms-field
entanglement and disentanglement through the dynamics
\cite{Gea1,Gea2,Phonx1,Phonx2,Phonx3,Buz1}. Finally the collapses
and revivals will be clearly connected to the evolution of the
$\textmd{Q}$-function \cite{Wolfgang,Vogel}. In this section we
shall discuss these quantities. We shall deal with three quantities
namely,
\begin{equation}\label{invt}
W_{T}(t)=\sum_{n=0}^{\infty }\left\{|X_{1}^{(n)}|^{2}-|X_{4}^{(n)}|^{2}\right\}\\
\end{equation}
\begin{eqnarray}\label{inva}
W_{a}(t)&=&\langle \Psi (t)|\hat{\sigma}_{z}^{(a)}|\Psi (t)\rangle
=\sum_{n=0}^{\infty
}\{|X_{1}^{(n)}|^{2}+\frac{1}{3}|X_{2}^{(n)}|^{2}\nonumber\\
&-& |X_{4}^{(n)}|^{2}- \frac{1}{3}|X_{3}^{(n)}|^{2}\}
\end{eqnarray}
\begin{equation}\label{invp}
P_{ini}(t)=| \langle \Psi (0)| \Psi (t)\rangle | ^{2}
\end{equation}

where, $W_{T}(t)$ is the total atomic inversion, $W_{a}(t)$ is the
inversion of the atom $a$ and $P_{ini}(t)$ is the probability of
occupation of the initial state. The two-atoms inversion is double
the one-atom inversion. In Fig.\ref{P1_inv} we show these quantities
for the atoms being initially (a) in the excited state $|
\Psi_{a}(0)\rangle=| eee \rangle$, or (b) the atoms initially in a
\textmd{GHZ}-state \cite{GHZ}, $| \Psi_{a}(0)\rangle=| \textmd{GHZ}
\rangle=\frac{1}{\sqrt{2}}(|ggg\rangle+|eee\rangle)$, which has the
property that tracing over any one qubit results in a maximally
mixed state containing no entanglement between the remaining two
qubits or (c) in genuine entangled \textmd{W}-state "Werner state"
\cite{W}, $|\Psi_{a}(0)\rangle=|\textmd{W}\rangle=\frac{1}{\sqrt 3}
(|eeg\rangle + |ege\rangle + |gee\rangle)$, for this state when
tracing over any one qubit the average remaining bipartite
entanglement is maximal. Before we dwell in the discussion of the
figures we present a rough analysis about the time dependent
quantities in the expressions (\ref{invt}, \ref{inva} and
\ref{invp}). Since we are dealing with a coherent state for which
$\bar{n}=100$, it is well known that the effective excitation will
be due to the photons within the range
$|n-\bar{n}|\leq\triangle{n}=\sqrt{\bar{n}}$. For these photons and
those within a reasonably range to them, the radicands appearing in
the expressions for $\beta$, $\gamma$ and $\eta$ in Eq.(\ref{Rabi})
can be approximated to $\sqrt{n}$ and the expressions for $\mu_{1}$,
$\mu_{2}$ are then given by "9n" and "n" respectively. Thus the Rabi
frequencies appearing in the expressions for ${\cal{U}}_{ij}$ are to
be approximated as $3\sqrt{n}$ and $\sqrt{n}$ respectively. Keeping
these points in mind, we find the following for the case of the
atomic initial state $| \Psi_{a}(0)\rangle=|eee\rangle$, i.e., the
three atoms are in their excited states: Regarding the quantity
$W_{T}(\tau)$, the time-dependent summand is proportional to
$\frac{1}{16}(\cos{6\sqrt{n}\tau}+15\cos{2\sqrt{n}\tau})$,
containing superharmonics which results in three revival times
$\frac{2\pi}{3}\sqrt{\bar{n}}$, $\frac{4\pi}{3}\sqrt{\bar{n}}$ and
$2\pi\sqrt{\bar{n}}$ related to the trigonometric functions with two
equal heights for the $1^{st}$ and $2^{nd}$ revival and 16 times
larger at the $3^{rd}$ revival. This appears clearly in
Fig.\ref{P1_inv}.(a1). On the other hand for the single atom
population inversion $W_{a}(t)$, the time dependent summand is
proportional to $\cos{2\sqrt{n}\tau}$, which results in a single
revival time at $2\pi\sqrt{\bar{n}}$ which is depicted in
Fig.\ref{P1_inv}.(a2). The $P_{ini}(\tau)$ has the term
$\frac{1}{32}(10+\cos{6\sqrt{n}\tau}+6\cos{4\sqrt{n}\tau}+15\cos{2\sqrt{n}\tau})$
in its summand, composed of 2 and 3 superharmoics giving rise to
revivals times at $\frac{2\pi}{3}\sqrt{\bar{n}}$,
$\pi\sqrt{\bar{n}}$, $\frac{4\pi}{3}\sqrt{\bar{n}}$ and
$2\pi\sqrt{\bar{n}}$. The second revival is larger than the $1^{st}$
and $3^{rd}$ because of the coefficients of $\cos{6\sqrt{n}\tau}$
and $\cos{4\sqrt{n}\tau}$ whiles the $4^{th}$, at
$2\pi\sqrt{\bar{n}}$, is the highest as the four revivals coincide.
This is shown faithfully in Fig.\ref{P1_inv}.(a3) whereas the
quantity $P_{ini}$ fluctuates
around $\frac{5}{16}$.\\

  Now we look at the initial state $|\Psi_{a}(0)\rangle=\frac{1}{\sqrt 2}(|ggg\rangle + |eee\rangle)$
Figs.\ref{P1_inv}(1b,2b,3b), and applying the same analysis as in
the previous case, we find the following:- The quantity
$W_{T}(\tau)$ as well as $W_{a}(\tau)$ is almost zero as can be seen
from Fig.\ref{P1_inv}(b1) and Fig.\ref{P1_inv}(b2). In contrast, the
quantity $P_{ini}(\tau)$ depends on the summand of the form
$\frac{1}{8}(5+\cos{4\sqrt{n}\tau})$ which results in revival times
$\pi\sqrt{\bar{n}}$ and $2\pi\sqrt{\bar{n}}$ shown clearly in
Fig.\ref{P1_inv}(b3). Also it exhibits the fluctuations around
$\frac{5}{8}$. These results show that, starting from the
$\textmd{GHZ}$-state would result in coherent trapping and the atoms
almost do not interact with the field. This is shown in the value of
$\frac{5}{8}$ for the probability for the atoms staying in the
initial state.\\

  The case of the atomic initial Werner state namely
$|\Psi_{a}(0)\rangle=\frac{1}{\sqrt 3}(|eeg\rangle + |ege\rangle +
|gee\rangle)$ is shown in Figs.\ref{P1_inv}(c1,c2,c3). Analysis of
the summand of $W_{T}(\tau)$ reveals that the time-dependence is of
the form $(\cos{6\sqrt{n}\tau}+\cos{2\sqrt{n}\tau})$. This amounts
to the revival times $\frac{2\pi}{3}\sqrt{\bar{n}}$,
$\frac{4\pi}{3}\sqrt{\bar{n}}$ and $2\pi\sqrt{\bar{n}}$ with the
amplitude at the $3^{rd}$ revival twice as much as the $1^{st}$
revival. Whereas the single atom population inversion has a single
revival at $2\pi\sqrt{\bar{n}}$ because the time dependent term in
the summand is proportional to $\cos{2\sqrt{n}\tau}$. It is
worthnoting that the fluctuations in the population inversion are
the highest for the $| eee \rangle$ initial state and the lowest for
the $\textmd{GHZ}$ initial state showing an almost coherent trapping
\cite{Zaheer} for the latter in such a way to resist exchange of
energy with the field, whereas the \textmd{W}-state is in the
middle, showing a modest degree of exchange of the energy with the
field.
\maketitle
\section{Cooperative and Atoms-Pairwise entanglement}
\begin{figure*}
\includegraphics[width=\textwidth]{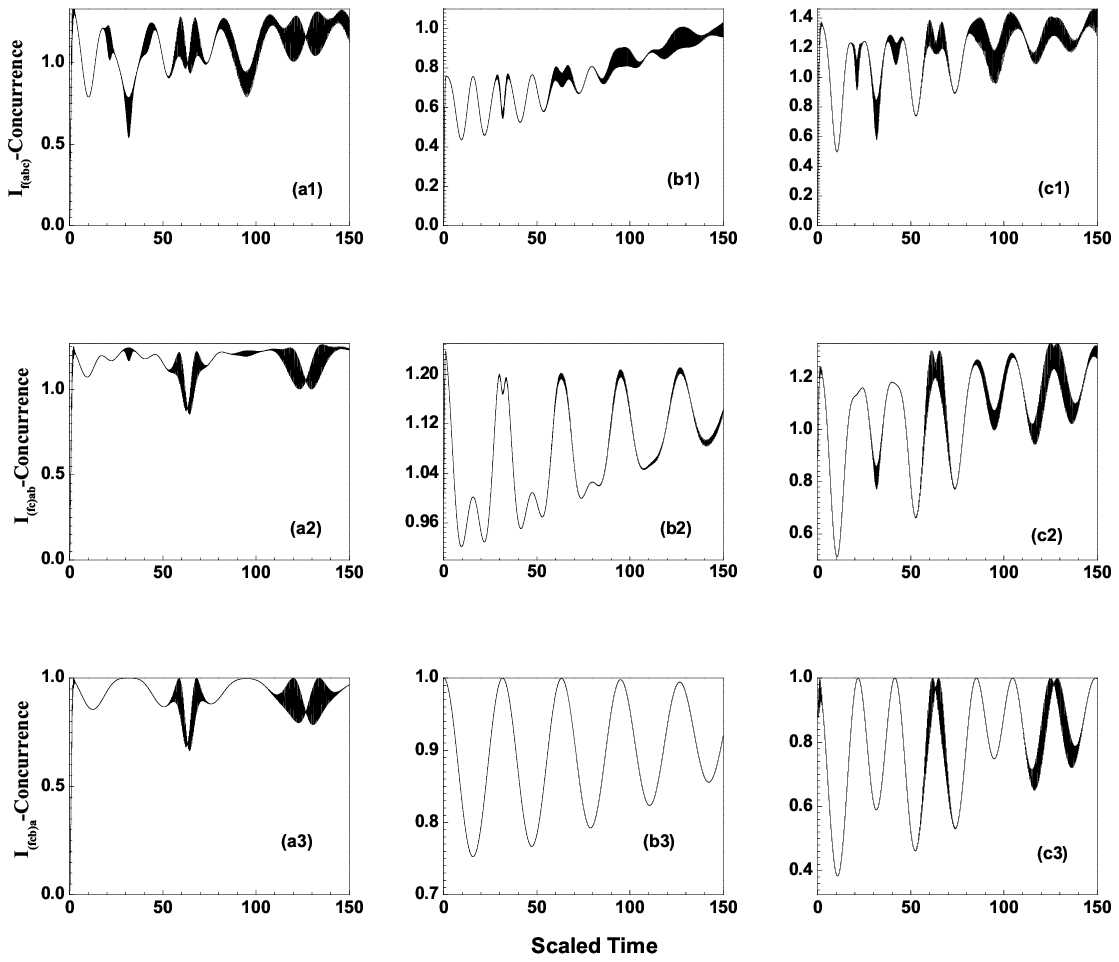}
\caption{The I-concurrence for the three bipartite partitions in
(i), (a)$|eee\rangle$-initial
state,(b)$|\textmd{GHZ}\rangle$-initial state and (c)
$|\textmd{W}\rangle$-initial state.}\label{Pur}
\end{figure*}
\begin{figure}[tb]
\begin{center}
\includegraphics[angle=0,scale =1]{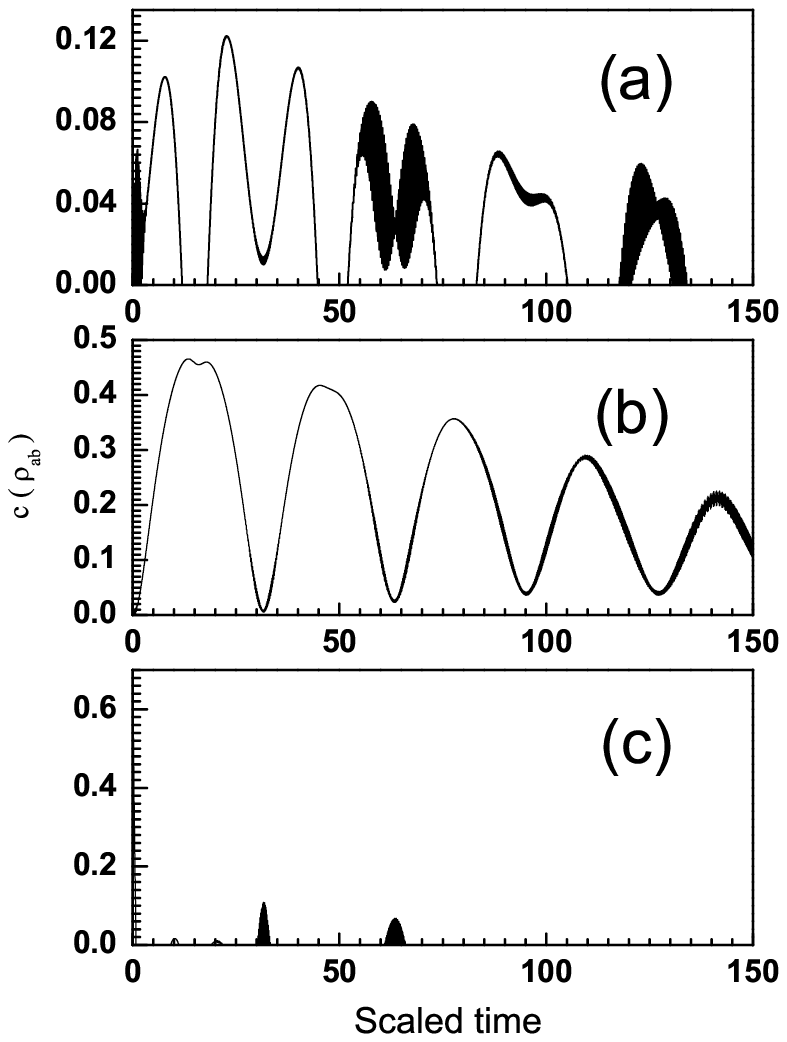}
\caption{Evolution of the atoms-pairwise entanglement, by the
concurrence of $\hat\rho_{ab}$ For (a)$|eee\rangle$-initial
state,(b)$|\textmd{GHZ}\rangle$-initial state and (c)
$|\textmd{W}\rangle$-initial state.}\label{TAN}
\end{center}
\end{figure}
 The system under investigation is a multipartite system initially in
an over all pure state, occupying the following Hilbert space ${\cal
H}={\cal H}_{a}\otimes{\cal H}_{b}\otimes{\cal H}_{c}\otimes{\cal
H}_{f}$ with dimension $2\otimes2\otimes2\otimes\infty$. Since we
consider here only the cooperative case, whereas the Hamiltonian is
symmetric under atom exchange, thus there are only eight
nonequivalent partitions of bipartite subsystems namely: (i)
$f\otimes(abc)$, $(fc)\otimes\,ab$, $(fcb)\otimes\,a$, (ii)
$f\otimes(ab)$, (iii) $f\otimes a$, (iv) $(fb)\otimes\,a$, (v)
$a\otimes(bc)$ and (vi) $a\otimes b$, corresponding to the field
times the whole atomic system, two-atoms times the field with the
remaining atom, one-atom times the field with the other two atoms,
the field times two-atoms, the field times one-atom, one atom times
the field with one-atom, one-atom times the other two atoms and
one-atom times one-atom respectively.\\

 These partitions except (i), are obtained by tracing over one or more
Hilbert-subspaces of the total Hilbert space and hence they are
generally mixed states. For convenience and simplicity we will only
study the entanglement evolution of the partitions (i) and the
entanglement of the last partition, i.e. atom-atom entanglement. In
the following section the entanglement of the bipartite partition
(v), will be discussed together with the phenomenon of entanglement
sharing \cite{Coffman,Deutsch} and the residual 3-particle
entanglement using the negativity.\\

  For the bipartite partitions in (i), each one of them start from an
over all pure state and a number of widely accepted measures of
entanglement are available. An easier way to quantify the
entanglement in this bipartite partitions is to use the square of
the pure-state I-concurrence introduced in Ref. \cite{Rungta},
\begin{equation}\label{Iconc}
C^{2}(\Psi)=2\,\nu_{d_{1}}\nu_{d_{2}}
[1-\textrm{Tr}\{\hat{\rho}_{f}^{2}\}]=2\,\nu_{d_{1}}\nu_{d_{2}}
[1-\textrm{Tr}\{\hat{\rho}_{abc}^{2}\}]
\end{equation}
This generalizes the original concurrence's notion introduced by
Hill and Wootters \cite{Hill2} $C(\Psi)\equiv \sqrt{\langle
\Psi|S_{2}\,\otimes\,S_{2}(|\Psi \rangle\langle
\Psi|)|\Psi\rangle}=|\langle
\Psi|\sigma_{y}\otimes\sigma_{y}|\Psi^{*}\rangle|$ for pairs of
qubits in a joint pure state $|\Psi\rangle$, to be applied for pairs
of quantum systems of arbitrary dimension $d_{1}\otimes d_{2}$ in a
joint pure state. The concurrence is defined with the help of a
superoperator $S_{2}$, whose action on a qubit density operator
$\rho=|\Psi\rangle\langle\Psi|$ is to flip the spin of the qubit
density operator $S_{2}(\rho) =\rho^{*} $;
$\rho^{*}=|\tilde{\Psi}\rangle\langle\tilde{\Psi}|$, where
$|\tilde{\Psi}\rangle=\sigma_{y}\otimes\sigma_{y}|\Psi^{*}\rangle$ ,
the asterisk denotes the complex conjugate and $\sigma_{y}$ is the
Pauli matrix. Rungta et al. \cite{Rungta} use the formalism for
superoperators \cite{Alickibook} to generalize the spin-flip
superoperator $S_{2}$ for a qubit to a superoperator $S_{d}$ that
acts on qudit states (d-dimensional states). For defining an
I-concurrence, one should choose the scaling factor $\nu_{d}$ to be
independent of $d$, otherwise, the pure state I-concurrence could be
changed simply by adding extra, unused dimensions to one of the
subsystems. To be consistent with the qubit concurrence, one should
choose $\nu_{d}=1$. With this choice the pure-state I-concurrence
runs from zero for product states to $I_{max}
=\sqrt{\frac{2(m-1)}{m}}$, where $m= min(d_{1},d_{2})$, for a
maximally entangled state. Henceforth we will use only the term
I-concurrence when refereing to it.\\

  On the other hand, Wootters extended the concurrence notation to the case of
a two qubits in an arbitrary joint mixed state, he showed that the
entanglement of formation \cite{Been3,Been4} of an arbitrary
two-qubits mixed state $\rho$ can be written in terms of the minimum
average pure-state concurrence of ensemble decompositions of $\rho$,
and he derived an explicit expression for this minimum in terms of
the eigenvalues of $\rho \tilde{\rho}$\cite{woot}.
\begin{equation}\label{tang}
C(\rho)=
\textmd{max}\{0,\lambda_{1}-\lambda_{2}-\lambda_{3}-\lambda_{4}\}
\end{equation}
where $\lambda_{i}$'s are the eigenvalues, in decreasing order, of
the Hermitian matrix
$\textmd{R}=\sqrt{\sqrt{\rho}\tilde{\rho}\sqrt{\rho}}$.
Alternatively, one can say that the $\lambda_{i}$'s are the square
roots of the eigenvalues of the matrix $\rho\,\tilde{\rho}$ and each
$\lambda_{i}$ is a non-negative real number. The spin-flipped state
$\tilde{\rho}$ is obtained by spin flipping, namely
$\tilde{\rho}=(\sigma_{y}\otimes\sigma_{y})\rho^{*}(\sigma_{y}\otimes\sigma_{y})$.
where again the asterisk denote the complex conjugate. For a pure
state $|\Psi\rangle$, R has only one eigenvalue that may be nonzero,
namely, $C(\Psi )$ Eq.(\ref{Iconc}). Wootters called this minimum
average, the concurrence of the mixed state. The entanglement of the
last partition, i.e. atom-atom entanglement can be investigated
using this formula.\\

  It is worthnoting that the remanning partitions are all in a mixed
state with dimensions $2\otimes \infty$, $4\otimes \infty$ and
$2\otimes4$, which can not be quantified by any of the above
mentioned entanglement measures. Following Tessier et al.
\cite{Deutsch}, we can quantify the entanglement for all the above
partitions using Osborne's formula \cite{Osbor} for the I-tangle
$\tau(\rho_{AB})$ for mixed states $\rho_{AB}$ of a pair
$\textmd{AB}$ of qudits, of dimensions $d_{A}$ and $d_{B}$, having
no more than two nonzero eigenvalues, i.e., $\rho_{AB}$ with rank no
greater than 2. The I-tangle $\tau$ between $\textmd{A}$ and
$\textmd{B}$ is given by the expression
\begin{equation}\label{Itang}
\tau(\rho_{AB})=
\textmd{Tr}\{\rho_{AB}\tilde{\rho}_{AB}\}+2\,\lambda_{min}(1-\textmd{Tr}\{\rho_{AB}^{2}\})
\end{equation}
where $\lambda_{min}$ is is the smallest eigenvalue of the real
symmetric $3\times3$ matrix $\textmd{M}$ as defined in \cite{Osbor}\\

  In Fig.\ref{Pur} we show the I-concurrence of the reduced density
matrices ${\rho}_{abc}$, ${\rho}_{ab}$ and ${\rho}_{a}$, (a) for the
$|eee\rangle$ case, (b) the $|\textmd{GHZ}\rangle$ case and (c) the
$|\textmd{W}\rangle$ case. This type of entanglement may be called a
cooperative entanglement. Figs.2(a1,a2,a3,b1,c1) begin from zero
which corresponds to the initial product state while
Figs.(b2,b3,c2,c3) start from 3-particle maximum entanglement state.
 They suddenly increase in the collapse region to reach the maximum
values, $I_{max}= 1.34, 1.23$ and $1$ for $f\otimes(abc)$,
$(fc)\otimes\,ab$ and $(fcb)\otimes\,a$ respectively. The exception
is the entanglement between the field and the total atomic system in
the case of the $\textmd{GHZ}$-state in Fig.\ref{Pur}(b1), which did
not reach this maximum. This reflects the robustness of the initial
entanglement of this state, i.e., the initial entanglement between
the atoms resists the interaction and the energy exchange between
them and the field as shown in the discussion of the population
inversion of last section. When this entanglement becomes rather
weak the field-atoms entanglement increase rapidly to the maximum
value $I_{max}$. On the other hand the $\textmd{W}$-state case did
not show such robustness. There are many local minima and maxima
asymptotically increasing to reach the maximum values, $I_{max}$. It
is noticed that the $\textmd{GHZ}$-state shows more periodicity than
the other two cases. Finally, it is clear that the local maxima and
minima of the $f\otimes(abc)$ and $(a)\otimes\,fbc$ figures are
opposite to each other. Fig.\ref{TAN} shows the concurrence of the
reduced density matrix $C(\rho_{ab})$ for the different initial
states. In contrast to the above cooperative entanglement, the
atom-atom entanglement asymptotically decreases and vanishes. The
initially excited atoms and the $\textmd{GHZ}$ case start from zero
while the $|\textmd{W}\rangle$ case starts from maximum entanglement
$C(\rho_{ab})=\frac{2}{3}$ and then there is a sudden death
\cite{death,birth,birth2} and sudden rebirth "anabiosis"
\cite{anabiosis}. This phenomenon appears in this case as a
bipartite subsystem where the third atom and the field are traced.
\maketitle
\section{Three-particle residual entanglement}
\begin{figure*}
\includegraphics[width=\textwidth]{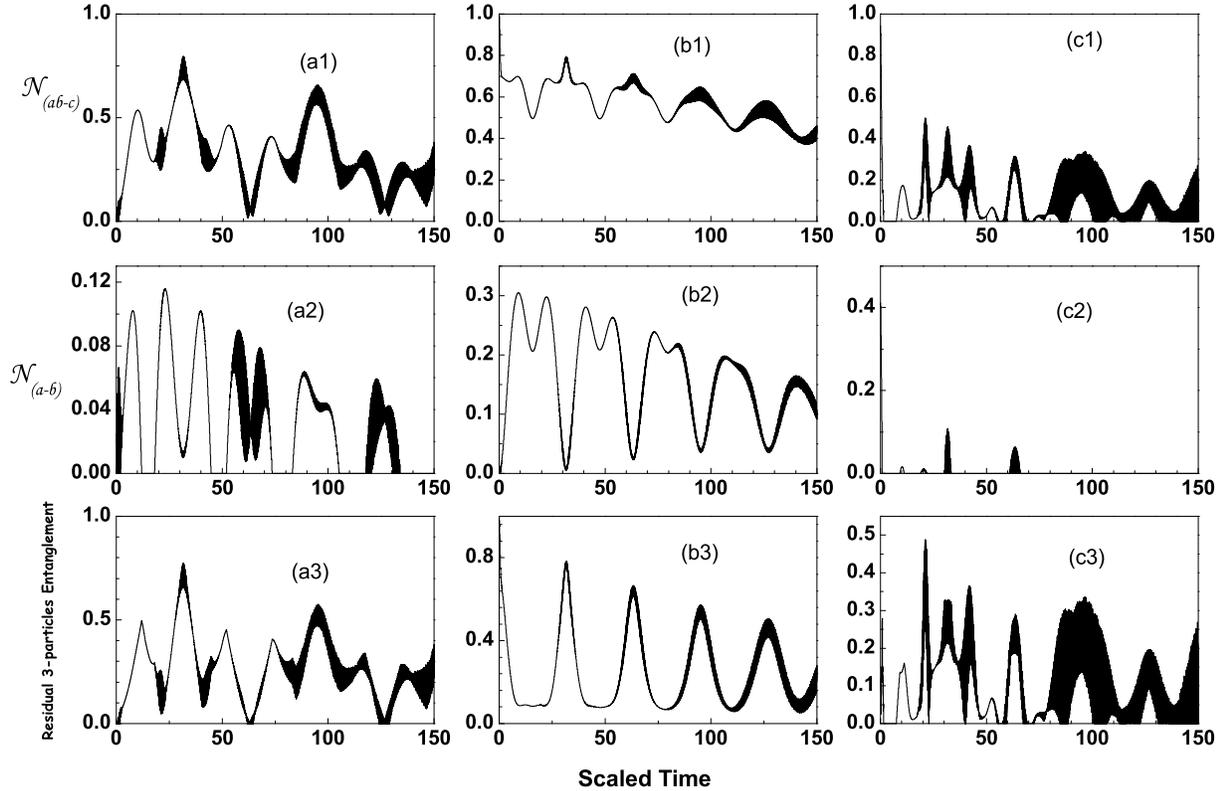}
\caption{Evolution of the negativity  ${\mathcal
N}_{a-bc}(\rho_{abc})$,  ${\mathcal N}_{a-b}(\rho_{ab})$ and the
residual 3-particles negativity ${\mathcal N}_{abc}$ for different
initial atomic states (a)$|eee\rangle$-initial
state,(b)$|\textmd{GHZ}\rangle$-initial state and (c)
$|\textmd{W}\rangle$-initial state.}\label{NEG}
\end{figure*}
  By using the notion of negativity \cite{zycz1,vidal}, we shall study the
phenomena of entanglement sharing and the residual 3-particle
entanglement "three-way entanglement" \cite{Coffman,Deutsch}.
Coffman et al \cite{Coffman} introduced the tangle notation to
quantify the entanglement of three qubits A, B, and C in a joint
pure state and they discussed entanglement sharing between three
particles. They found that unlike classical correlation, quantum
entanglement cannot be freely shared among the particles. More
precisely they found that the following inequality

\begin{equation}\label{threewayt1}
\tau_{A(BC)}\geq \tau_{AB}+\tau_{AC}
\end{equation}
holds for any three qubits in a joint pure state. Here the tangle
$\tau_{A(BC)}$ is the I-concurrence of Eq.\ref{Iconc}, between A and
the other two qubits as one entity, while the other two tangles
$\tau_{AB}$ and $\tau_{AC}$ are the squares of concurrence
Eq.\ref{tang}, between A and B, and A and C respectively.
Consequently they defined the 3-tangle (or residual tangle)
$\tau_{ABC}$ as follows:
\begin{equation}\label{threewayt}
\tau_{A(BC)}=\tau_{AB}+\tau_{AC}+\tau_{ABC}
\end{equation}
 To be read, the entanglement of A with the rest of the system is
equal to the entanglement of A with B alone, plus the entanglement
of A with C alone plus the 3-tangle of the whole system. The
3-tangle quantify the residual three-particle entanglement, which
cannot be accounted for by the pairwise entanglement. For the
\textmd{GHZ}-state we have $1 = 0 + 0 + \tau_{ABC}$. Also they
showed that the inequality (\ref{threewayt1}) is true in case of
three particle in a joint mixed
state\\

  On the other hand the Peres-Horodecki criterion for separability
\cite{peres,horo0} leads to a natural computable measure of
entanglement, called negativity \cite{zycz1,zycz2,vidal}. The
negativity is based on the trace norm of the partial transpose
$\rho^{T_{A}}$ of the bipartite mixed state $\rho_{AB}$, and
measures the degree to which $\rho^{T_{A}}$ fails to be positive.
The density matrices which represent physical systems are
non-negative matrices with unit trace, the partial transpose also
satisfies $\textmd{Tr}\{\rho^{T_{A}}\}=1$, but since it may have
negative eigenvalues $\mu_{i}$" for entangled states" therefore its
trace norm reads in general
\begin{equation}\label{trac}
\parallel \rho^{T_1}\parallel_1=1+2|\sum_{i} \mu_{i}|\equiv1+{\mathcal N}(\rho)
\end{equation}
where ${\mathcal N}(\rho)$ is the negativity, i.e. the absolute
value of twice the sum of the negative eigenvalues. Vidal and Werner
\cite{vidal} proved that the negativity ${\mathcal N}(\rho)$ is an
entanglement monotone and therefore it is a good measure of
entanglement. Following \cite{vidal}, the residual three-particle
entanglement ${\mathcal N}_{abc}$ in our case can be quantified
using the relation
\begin{equation}\label{threewayn}
{\mathcal N}_{abc}={\mathcal N}_{a-bc}(\rho_{abc})-{\mathcal
N}_{a-b}(\rho_{ab})-{\mathcal N}_{a-c}(\rho_{ac})
\end{equation}
\begin{eqnarray}\label{threewayn2}
{\mathcal N}_{a-bc}(\rho_{abc})&=&\parallel
\rho_{abc}^{T_a}\parallel_1-1\\
{\mathcal N}_{a-b}(\rho_{ab})&=&\parallel
\rho_{ab}^{T_a}\parallel_1-1\equiv\parallel
\rho_{ab}^{T_b}\parallel_1-1\\
{\mathcal N}_{a-c}(\rho_{ac})&=&\parallel
\rho_{ac}^{T_a}\parallel_1-1\equiv\parallel
\rho_{ac}^{T_c}\parallel_1-1
\end{eqnarray}
 The term ${\mathcal N}_{a-bc}(\rho_{abc})$ quantifies the
strength of quantum correlations between the atom "$a$" and the
other two atoms. The remanning two terms quantify the pairwise
entanglement between the atom "$a$" and $b$ or $c$. Note that the
partial trace operation belongs to the set of local operations and
classical communication (LOCC) under which the entanglement can not
increase. Therefore the left hand side of Eq.17 is a residual
three-particle entanglement. In our case ${\mathcal
N}_{a-b}(\rho_{ab})={\mathcal N}_{a-c}(\rho_{ac})$ because of the
symmetry. In Fig.\ref{NEG} we show the results for the different
initial states. We see that the residual 3-particle entanglement
changes between local maxima and minima during the evolution. The
local maxima occur at half-revival times, when the the atoms became
most disentangled from the field. Asymptotically this residual
entanglement vanishes and the atomic system becomes maximally
entangled with the field. From Figs. \ref{TAN} and \ref{NEG} it is
shown that the negativity measure of two-atom entanglement is less
than or at most equal to the concurrence as it is conjectured in
\cite{zycz2}.
\maketitle
\section{Q-function}
\begin{figure*}
\includegraphics[angle=0,scale =1.0]{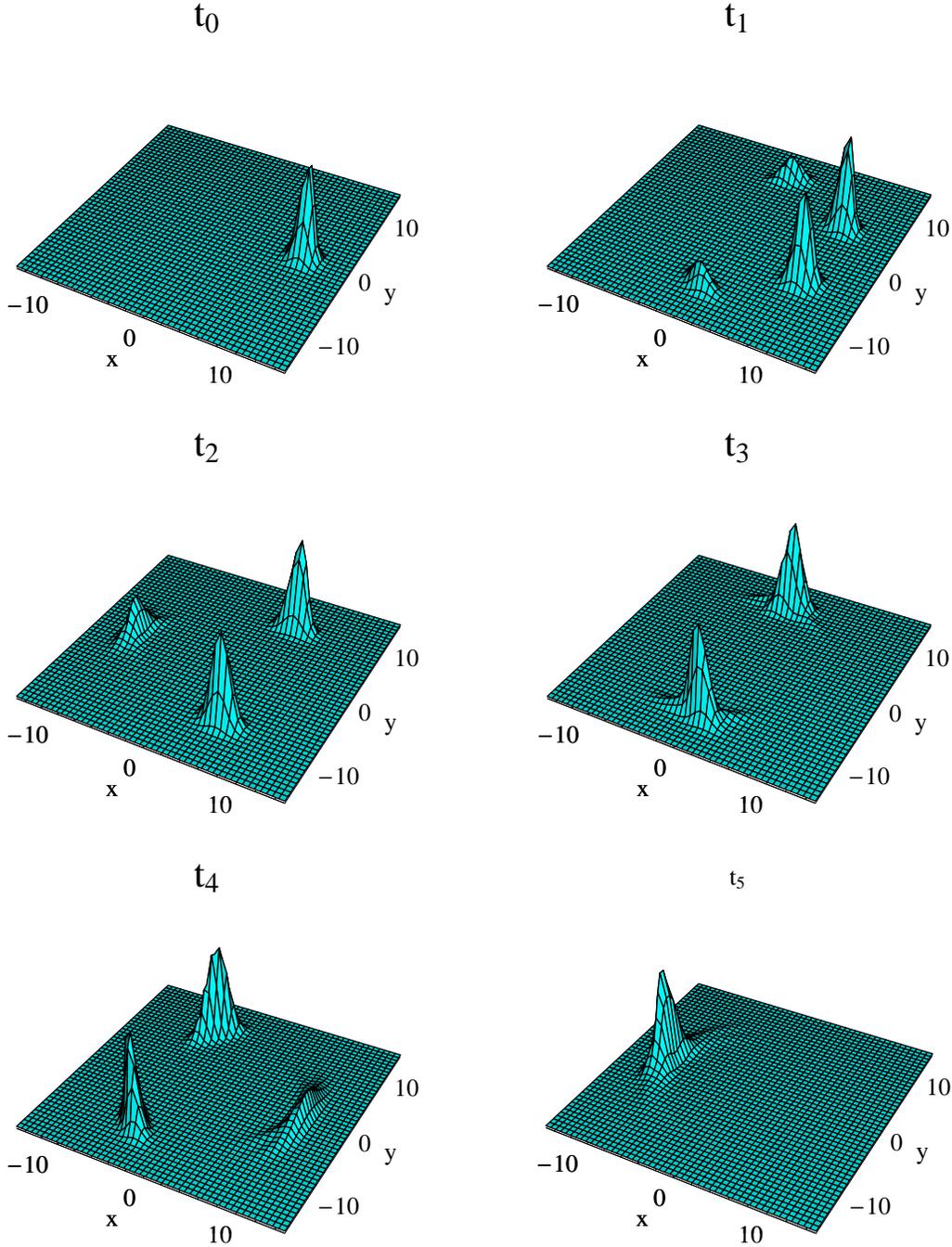}
\caption{The Q-function for the initial atomic
state$\ket{\Psi(0)}=\ket{eee}$}\label{Qfune}
\end{figure*}
\begin{figure*}
\includegraphics[angle=0,scale =1.0]{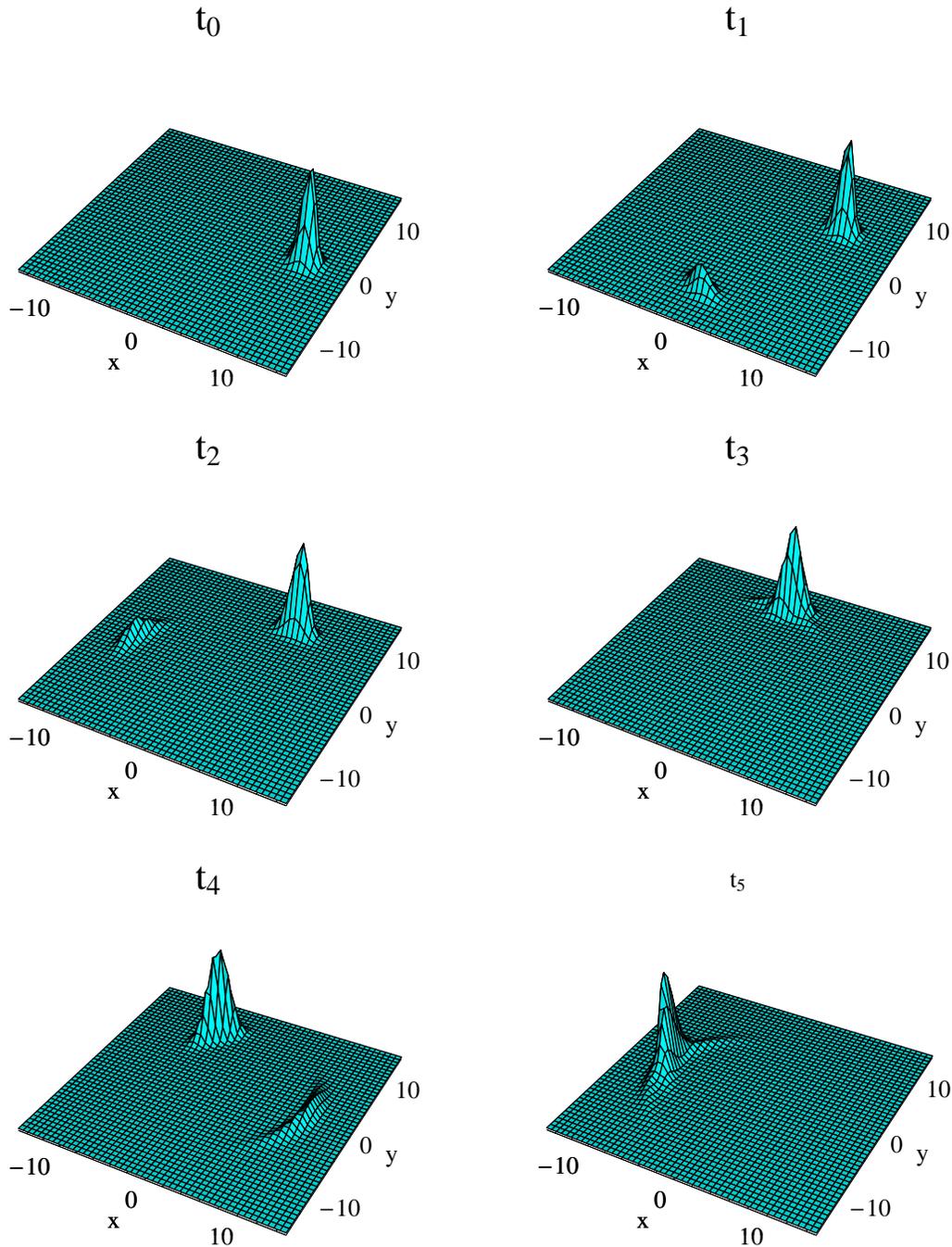}
\caption{The Q-function for the initial atomic
state$\ket{\Psi(0)}=\frac{1}{\sqrt{2}}(\ket{eee}+\ket{ggg})$}\label{Qfuns}
\end{figure*}
\begin{figure*}
\includegraphics[angle=0,scale =1.0]{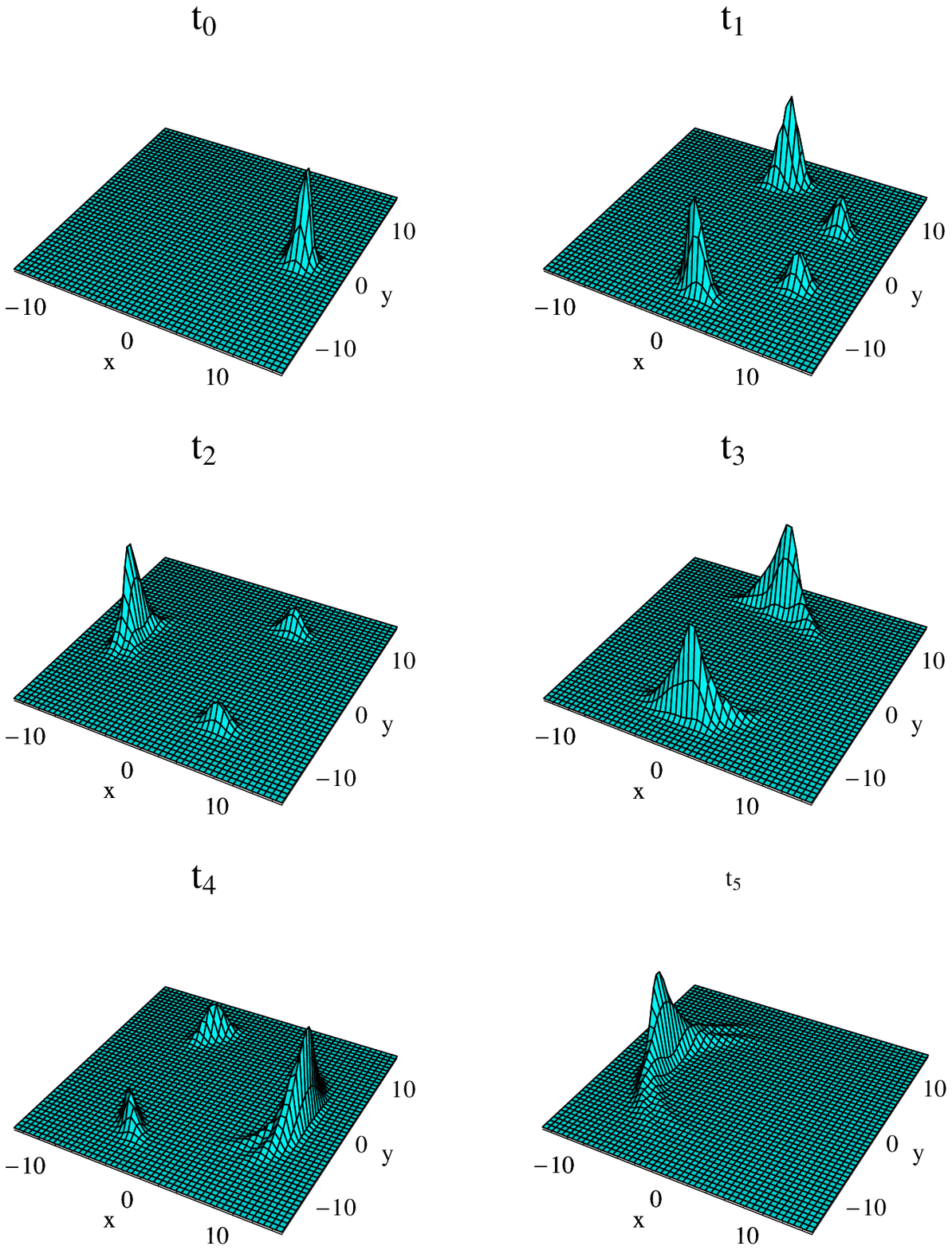}
\caption{The Q-function for the initial atomic
state$\ket{\Psi(0)}=\frac{1}{\sqrt{3}}
(\ket{eeg}+\ket{ege}+\ket{gee})$}\label{Qfunw}
\end{figure*}
In this section we return to the entanglement between the field and
the atomic system but through the field dynamic, namely we will
discuss the field quasi-probability distribution function, the
$\textmd{Q}$-function \cite{scully}, defined as
$Q(\beta,t)=\frac{1}{\pi }\langle \beta |\rho_{f}\left( t\right)
|\beta \rangle$. By using Eq.(\ref{Amp}) we get the following form
of the $\textmd{Q}$-function in terms of probability amplitudes.
\begin{equation}\label{Qfun}
Q(\alpha,t)=\frac{e^{-|\beta|^{2}}}{\pi
}\sum_{i=1}^{4}\left|\sum_{n=0}^{\infty}\frac{(\beta^{*})^{n+i-1}}{\sqrt{(n+i-1)!}}X_{i}^{(n)}
\right|^{2}
\end{equation}
Figs.\ref{Qfune},\ref{Qfuns}, \ref{Qfunw} show the $Q$-function for
different initial atomic states at different characteristic times
$t_{0}=0$, $t_{1}=\frac{\pi}{3}\sqrt{\bar{n}}$,
$t_{2}=\frac{2\pi}{3}\sqrt{\bar{n}}$, $t_{3}=\pi\sqrt{\bar{n}}$,
$t_{4}=\frac{4\pi}{3}\sqrt{\bar{n}}$, $t_{5}=2\,\pi\sqrt{\bar{n}}$.
The field is initially in a coherent state with average photon
number $|\alpha_{0}|^{2}=\bar{n}=100$. At $\tau=0$ the shape of the
$\textmd{Q}$-function is a Gaussian centered at $(\sqrt{\bar{n}},0)$
as expected for the initial coherent state. But as time develops we
found different types of behavior associated with each initial
atomic state. For the state $|eee\rangle$ depicted in
Fig.\ref{Qfune}; we note that the single peak is splitted into two
pairs, a pair with large amplitude moving slowly in opposite
directions and the other pair of the smaller amplitudes moving
fastly in opposite directions. To shed some light on this behavior,
rough analysis of formula (\ref{Qfun}), similar to that used in the
atomic inversion, shows that time dependent factors in the summand
behave like
$\{\cos3(\sqrt{n}-\sqrt{m})\tau+3\cos(\sqrt{n}-\sqrt{m})\tau\}$.
This form exhibits clearly the appearance of the pairs and the first
term represents the faster pair while the second term represents the
slower pair with amplitude 3 times larger than the faster ones. Note
the appearance of the Schr\"{o}dinger cat-state \cite{Buz1}, at
half-revival ($\tau=t_{3}$). The case of the state
$\frac{1}{\sqrt{2}}(|ggg\rangle+|eee\rangle)$ is quite interesting.
Here the peak splits into only two peaks; a small faster peak and a
large slower peak. The analysis shows that the time-dependent factor
in the summand behaves as
$\{e^{3i(\sqrt{m}-\sqrt{n})\tau}+3e^{i(\sqrt{n}-\sqrt{m})\tau}\}$
which exhibits the two single peaks, they move in opposite
directions and the larger is the slower peak. The third case of the
Werner state as depicted in Fig.\ref{Qfunw} shows an opposite
behaviour to that of the $|eee\rangle$ state. The pair with the
larger amplitude moves faster than the pair with smaller amplitude.
The time dependent-factor in the summand in this case behaves like
$\{3\cos3(\sqrt{n}-\sqrt{m})\tau+\cos(\sqrt{n}-\sqrt{m})\tau\}$
which demonstrates the fast movement of the larger components and
the slow movement of the smaller components.  It is to be remarked
that at half revival time starting from either the uppermost excited
state ($|eee\rangle$) or the Werener state the Schr\"{o}dinger
cat-state is produced. On the other hand starting from the
\textmd{GHZ}-state we find that at half revival time the field state
returns to its original state. Because as the larger component moves
$\frac{\pi}{2}$ in the phase space the smaller and faster component
would move $\frac{3\pi}{2}$ on the other direction and meets the
larger one. The previous behavior is also demonstrated in
Fig.\ref{P1_inv}.(b) regarding the atomic inversions\\

   Now looking at the various entanglement evolution
Figs.\ref{Pur} and \ref{NEG} we find that there is a clear
connection between the $\textmd{Q}$-function dynamic and various
entanglement evolutions. For the $|eee\rangle$ and
$|\textmd{W}\rangle$ initial states, at half the first-revival time
$t_{1}$ which correspond to a local minima of I-Concurrence,
$I_{f(abc)}$ in the $|eee\rangle$ case and to the minimum value in
the $|\textmd{W}\rangle$ case, the faster peaks became most apart
from each other. Also at that time the residual 3-particle
entanglement has a local maximum. At the first-revival time $t_{3}$
these two peaks collide and recombine, and this corresponds to a
local minima of $I_{f(abc)}$ in the two cases, a local minima of the
residual 3-particle in $|eee\rangle$ case and a maximum value in the
$|\textmd{W}\rangle$ case. Also we note that the amplitude of these
two peaks increase significantly in the $|\textmd{W}\rangle$ case.
This behavior can be connected to Figs.\ref{TAN}.(c1),(c2),(c3),
where it is clear that at that time $t_{4}$ the entanglement of all
the atomic ensembles with the field is minimum and this is not the
case in the $|eee\rangle$ state. The last interesting feature we
want to mention here for the $|eee\rangle$ and Werener cases is
that, at $t_{5}$, the Q-function, shows phase squeezing.

\maketitle
\section{Discussion and Conclusion}
In this paper we consider a system of identical three two-level
atoms interacting at resonance with a single-mode of the quantized
field in a lossless cavity. The initial cavity field is prepared in
the coherent state while the atoms are taken in different initial
states, namely the atoms taken to be in the excited state,
$|eee\rangle$, the $\textmd{GHZ}$ entangled state and the Werener
entangled state. For this system we investigated different kinds of
entanglement, atoms-field cooperative and atoms-pairwise
entanglements. We use the concurrence, the generalized I-concurrence
and the negativity as measures of these types of entanglements. The
relationship between this entanglement and the collapse and revival
in the atomic inversion is investigated. Also the $Q$-functions for
different cases are discussed and connected to different
entanglements evolutions of the system. Most noteworthy we found
that the $\textmd{GHZ}$-state is more robust against energy losses,
and showing almost coherent trapping. Also one can say that the
entanglement of $\textmd{GHZ}$-state is more robust than the
$\textmd{W}$-state. These two different behaviors have been
distinctly shown through the study, perfectly depicted in
Fig.\ref{P1_inv}.(b1) and Fig.\ref{NEG}.(b1) respectively. This
suggests that the $\textmd{GHZ}$-state may show more resistance to
the decoherence phenomena than any other three-partite entanglement.
Consequently it may be a primary candidate for many quantum
information tasks, which need a three-partite entanglement state. In
fact the $\textmd{GHZ}$-state is a resource for many applications,
these include, quantum secret sharing \cite{200}, open destination
teleportation \cite{201} and quantum computation \cite{202}. On the
other hand for the other two cases ($|eee\rangle$ and $\textmd{W}$
states) the Schr\"{o}inger cat-state is produced. Another
interesting feature is the clear link between the
\textmd{Q}-function dynamics and various entanglement evolution.
Finally we found that, while the $\textmd{W}$-state, is the state
with the maximal possible bipartite entanglement in the reduced
two-qubit states, its initial entanglement vanishes very rabidly.
Moreover the production of such pairwise entanglement through the
evolution is very small. This in contrast to the other two cases
which have no pairwise entanglement initially but such entanglement
increase greatly through the evolution. Sudden death and sudden
revival of atoms-pairwise entanglement are produced with the
$\textmd{W}$-state

\newpage
\maketitle
\appendix
\section{Appendix : Evolution matrix for the probabilities amplitudes}\label{sec:u}
  In this appendix we give the explicit form of the elements of the
evolution matrix ${\cal U}$ appearing in Eq.(\ref{evm})
\begin{widetext}
\begin{eqnarray}
{\cal U}_{11}&=&\left[\frac{\mu_{1}-4\beta^{2}-3\eta^{2}}
{(\mu_{1}-\mu_{2})}\cos(\sqrt{\mu_{1}}\,\tau)+\frac{\mu_{2}-4\beta^{2}-3\eta^{2}}
{(\mu_{2}-\mu_{1})}\cos(\sqrt{\mu_{2}}\,\tau)\right]
\nonumber \\
{\cal U}_{12}&=&-i\sqrt{3}\,\gamma\,\left[\frac{\mu_{1}-3\eta^{2}}
{\sqrt{\mu_{1}}(\mu_{1}-\mu_{2})}\,\sin(\sqrt{\mu_{1}}\,\tau)
+\frac{\mu_{2}-3\eta^{2}}
{\sqrt{\mu_{2}}(\mu_{2}-\mu_{1})}\,\sin(\sqrt{\mu_{2}}\,\tau)\right]
\nonumber \\
{\cal
U}_{13}&=&2\sqrt{3}\beta\gamma\left[\frac{1}{(\mu_{1}-\mu_{2})}
\cos(\sqrt{\mu_{1}}\,\tau) +\frac{1}
{(\mu_{2}-\mu_{1})}\cos(\sqrt{\mu_{2}}\,\tau)\right]\nonumber\\
{\cal U}_{14}&=&-6i\beta\gamma\eta\left[\frac{1}
{\sqrt{\mu_{1}}(\mu_{1}-\mu_{2})}\sin(\sqrt{\mu_{1}}\,\tau)
+\frac{1}{\sqrt{\mu_{2}}(\mu_{2}-\mu_{1})}
\sin(\sqrt{\mu_{2}}\,\tau)\right]\nonumber\\
{\cal U}_{22}&=&\left[\frac{\mu_{1}-3\eta^{2}}
{(\mu_{1}-\mu_{2})}\cos(\sqrt{\mu_{1}}\,\tau)
+\frac{\mu_{2}-3\eta^{2}}
{(\mu_{2}-\mu_{1})}\cos(\sqrt{\mu_{2}}\,\tau)\right]\nonumber\\
{\cal U}_{23}&=&-2i\beta\left[\frac{\sqrt{\mu_{1}}}
{(\mu_{1}-\mu_{2})}\sin(\sqrt{\mu_{1}}\,\tau)
+\frac{\sqrt{\mu_{2}}}{(\mu_{2}-\mu_{1})}
\sin(\sqrt{\mu_{2}}\,\tau)\right]\nonumber\\
{\cal U}_{24}&=&2\sqrt{3}\beta\eta\left[\frac{1}{(\mu_{1}-\mu_{2})}
\cos(\sqrt{\mu_{1}}\,\tau)+\frac{1}
{(\mu_{2}-\mu_{1})}\cos(\sqrt{\mu_{2}}\,\tau)\right]\nonumber\\
{\cal U}_{33}&=&\left[\frac{\mu_{1}-3\gamma^{2}}
{(\mu_{1}-\mu_{2})}\cos(\sqrt{\mu_{1}}\,\tau)
+\frac{\mu_{2}-3\gamma^{2}}
{(\mu_{2}-\mu_{1})}\cos(\sqrt{\mu_{2}}\,\tau)\right]\nonumber\\
{\cal U}_{34}&=&-i\sqrt{3}\,\eta\,\left[\frac{\mu_{1}-3\gamma^{2}}
{\sqrt{\mu_{1}}(\mu_{1}-\mu_{2})}\,\sin(\sqrt{\mu_{1}}\,\tau)
+\frac{\mu_{2}-3\gamma^{2}}
{\sqrt{\mu_{2}}(\mu_{2}-\mu_{1})}\,\sin(\sqrt{\mu_{2}}\,\tau)\right]
\nonumber\\
{\cal U}_{44}&=&\left[\frac{\mu_{1}-4\beta^{2}-3\gamma^{2}}
{(\mu_{1}-\mu_{2})}\cos(\sqrt{\mu_{1}}\,\tau)
+\frac{\mu_{2}-4\beta^{2}-3\gamma^{2}}
{(\mu_{2}-\mu_{1})}\cos(\sqrt{\mu_{2}}\,\tau)\right]\nonumber
\end{eqnarray}
\end{widetext}

\end{document}